\documentclass[twocolumn,preprintnumbers,amsmath,amssymb,superscriptaddress]{revtex4}

\usepackage{graphicx}
\usepackage{dcolumn}
\usepackage{bm}
\usepackage{soul}
\usepackage{color}
\usepackage{epstopdf}
\usepackage[version=3]{mhchem}
\usepackage{lipsum}
\usepackage[outercaption]{sidecap}
\usepackage{floatrow}
\begin{document}
\title{Determination of Young's modulus of active pharmaceutical ingredients by relaxation dynamics at elevated pressures}
\author{Anh D. Phan}
\affiliation{Faculty of Materials Science and Engineering, Computer Science, Artificial Intelligence Laboratory, Phenikaa Institute for Advanced Study, Phenikaa University, Hanoi 12116, Vietnam}
\email{anh.phanduc@phenikaa-uni.edu.vn}

\date{\today}

\date{\today}

\begin{abstract}
A new approach is theoretically proposed to study the glass transition of active pharmaceutical ingredients and a glass-forming anisotropic molecular liquid at high pressures. We describe amorphous materials as a fluid of hard spheres. Effects of nearest neighbor interactions and cooperative motions of particles on glassy dynamics are quantified through a local and collective elastic barrier calculated using the Elastically Collective Nonlinear Langevin Equation theory. Inserting two barriers into Kramer's theory gives structural relaxation time. Then, we formulate a new mapping based on the thermal expansion process under pressure to intercorrelate particle density, temperature, and pressure. This analysis allows us to determine the pressure and temperature dependence of alpha relaxation. From this, we estimate an effective elastic modulus of amorphous materials and capture effects of conformation on the relaxation process. Remarkably, our theoretical results agree well with experiments.
\end{abstract}

\maketitle
\section{Introduction}
Amorphous drugs have been intensively investigated in recent years because they are more highly water-soluble and bioavailable than their crystalline counterparts \cite{49,50,51,52,53}. The structural disorders are achieved by cooling molten materials at a sufficient rapid rate to fall out of equilibrium. The non-equilibrium state leads to long-range molecular disordering and makes them liquid, particularly at high temperatures, while rheological properties behave as if solid. However, the main drawback of amorphous pharmaceuticals is that the thermodynamics becomes unstable and the structure can be recrystallized during manufacturing or storage process \cite{49,50}. To overcome these issues, molecular dynamics of amorphous drugs \cite{49,50,51,52,53} characterized by structural relaxation process is a crucial key. The relaxation is significantly slowed down with cooling and/or compressing \cite{63,64,65,66}. Thus, it is essential for industrial applications and fundamental science to gain insights into the pressure and temperature dependence of the relaxation time.

The Elastically Collective Nonlinear Langevin Equation (ECNLE) theory has been used to study molecular dynamics in supercooled and glassy state of various amorphous materials \cite{2,7,10,6,35,42,11,61,62}. The first version of the ECNLE theory proposed by Mirigian and Schweizer has successfully described the glass transition of polymers, thermal liquids, and colloidal systems \cite{6,7,10} at ambient pressure. This approach uses a hard-sphere fluid to describe glass-forming liquids and determine the structural relaxation time as a function of particle density. A predicted timescale ranges from 1 ps to $10^3$ s, which is far beyond the simulation timescale ($\leq 10^6$ ps). To map from density to temperature (a thermal mapping process), the authors use experimental equation-of-state data. Their first effort toward capturing compression effects \cite{10} accurately predicts variation of the dynamic fragility and glass transition temperature. However, deviations between theoretical and experimental results are quantitatively large. As missing experimental equation-of-state data, this method cannot be used. Very recently, we introduced a new thermal mapping based on the thermal expansion process to address this limitation \cite{42,11,35,61,62}. Now, the ECNLE theory can calculate the relaxation time of amorphous drugs, supramolecules, and multi-component systems at atmospheric pressure.

In recent works \cite{35,61}, we developed the ECNLE theory to consider compression effects on the structural relaxation process of amorphous drugs and supramolecules. A pressure-induced restriction of motion of a single molecule is modeled as a mechanical work, which modifies its dynamic free energy constructed based on nearest neighbor interactions. The dynamic constraint hinders cooperative motions and leads to the slowing down of molecular dynamics with increasing compression. Here, we use our new thermal mapping at ambient pressure and suppose that it does not change with compression. Although the approach has provided a quantitatively good description for a variety of experiments \cite{35,61}, there are several problems: (i) ignorance of physical intercorrelations among density, temperature, and pressure, (ii) the occupied volume equal to zero at 0 $K$ under isobaric condition, and (iii) the particle size is an adjustable parameter when determining the pressure dependence of structural relaxation time in unit of MPa or GPa. The latter problem limits a range of external pressure in theoretical calculations.

In this paper, we extend the ECNLE theory in a different way to investigate effects of pressure and molecular conformation on the glass transition of active pharmaceutical ingredients and glass-forming liquids. We connect the packing fraction of a hard-sphere fluid to pressure and elastic modulus that establishes an equation of state. The pressure-induced density change alters the static structural factor or radial distribution function. Thus, information of the pressure dependence is encoded in the dynamic free energy without using the hypothetical mechanical work as previous studies \cite{35,61}. This allows us to calculate the pressure and temperature dependence of the dynamic fragility and structural relaxation time for five materials. Our numerical calculations are quantitatively compared to experimental data. We also show how the elastic modulus is determined using broadband dielectric spectroscopy data.

\section{Theoretical background}
As a background, we briefly review the ECNLE theory. A hard-sphere fluid describing glass-forming liquids in ECNLE theory is characterized by a particle density, $\rho$, and a particle diameter, $d$ \cite{2,7,10,6,35,42,11,61,62}. By using the Persus-Yevick theory \cite{1}, we calculate the static structure factor, $S(q)$, and the radial distribution function, $g(r)$. Molecular dynamics of a tagged particle is mainly governed by interactions with its nearest neighbors, random white noise, and a friction force. These forces obey the nonlinear stochastic equation. Solving this equation associated with the force-force correlation function gives the free dynamic energy to quantify the nearest neighbor interaction
\begin{eqnarray}
\frac{F_{dyn}(r)}{k_BT} &=& \int_0^{\infty} dq\frac{ q^2d^3 \left[S(q)-1\right]^2}{12\pi\Phi\left[1+S(q)\right]}\exp\left[-\frac{q^2r^2(S(q)+1)}{6S(q)}\right]
\nonumber\\ &-&3\ln\frac{r}{d},
\label{eq:2}
\end{eqnarray}
where $k_B$ is Boltzmann constant, $T$ is ambient temperature, $r$ is the displacement, $q$ is the wavevector, and $\Phi = \rho\pi d^3/6$ is the volume fraction. The first term is responsible for the caging constraint, while the second term corresponds to the ideal fluid state.

This free energy profile provides physical quantities to describe local dynamics. At a dense system, the interparticle separation is sufficiently reduced to form a particle cage to dynamically confine a central particle. The particle cage radius is determined as a position of the first minimum of $g(r)$. The onset of transient localization occurs when a local barrier emerges in $F_{dyn}(r)$. The dynamic free energy has a local minimum and maximum position located at $r_L$ and $r_B$, respectively. From these, one calculates a barrier height, $F_B=F_{dyn}(r_B)-F_{dyn}(r_L)$, and a jump distance, $\Delta r =r_B-r_L$.

Diffusion of a particle from its trapping cage demands collective rearrangement of surrounding particles. The cooperative motions generate a displacement field, $u(r)$, which is nucleated from the cage surface and radially propagates through the remaining space. Assume that a dilation at the cage surface is small, the displacement field can be analytically calculated via Lifshitz's linear continuum mechanics \cite{5} and this is
\begin{eqnarray} 
u(r)=\frac{\Delta r_{eff}r_{cage}^2}{r^2}, \quad {r\geq r_{cage}},
\label{eq:3}
\end{eqnarray}
where $\Delta r_{eff}$ is the amplitude of the displacement field, which is \cite{6,7}
\begin{eqnarray} 
\Delta r_{eff} = \frac{3}{r_{cage}^3}\left[\frac{r_{cage}^2\Delta r^2}{32} - \frac{r_{cage}\Delta r^3}{192} + \frac{\Delta r^4}{3072} \right].
\end{eqnarray}

Because of the small displacement field, one can treat collective particles as harmonic oscillators having a spring constant $K_0 = \left|\partial^2 F_{dyn}(r)/\partial r^2\right|_{r=r_L}$ and the amplitude of oscillation $u(r)$. A sum of elastic potential energy of all oscillators known as the elastic barrier quantifies effects of collective motions on the structural relaxation  \cite{2,7,10,6,35,42,11,61,62}. This elastic barrier is
\begin{eqnarray} 
F_{e} = 4\pi\rho\int_{r_{cage}}^{\infty}dr r^2 g(r)K_0\frac{u^2(r)}{2}. 
\label{eq:5}
\end{eqnarray}

Equation (\ref{eq:5}) indicates that the nature of local and elastic barrier are distinct but have strong correlation. According to Kramer's theory, one can calculate the structural relaxation time 
\begin{eqnarray}
\frac{\tau_\alpha}{\tau_s} = 1+ \frac{2\pi}{\sqrt{K_0K_B}}\frac{k_BT}{d^2}\exp\left(\frac{F_B+F_e}{k_BT} \right),
\label{eq:6}
\end{eqnarray}
where $\tau_s$ is a short time scale and $K_B$=$\left|\partial^2 F_{dyn}(r)/\partial r^2\right|_{r=r_B}$ is the absolute curvature at the barrier position. $\tau_s$ is analytically obtained by \cite{2,7,10,6,35,42,11,61,62}
\begin{eqnarray}
\tau_s=g^2(d)\tau_E\left[1+\frac{1}{36\pi\Phi}\int_0^{\infty}dq\frac{q^2(S(q)-1)^2}{S(q)+b(q)} \right],
\label{eq:8}
\end{eqnarray}
where $b(q)=1/\left[1-j_0(q)+2j_2(q)\right]$, $j_n(x)$ is the spherical Bessel function of order $n$, and the Enskog time scale is $\tau_E \approx 10^{-13}$ s \cite{2,6,7,35,42,11,61,62}.

To compare ECNLE calculations with experiments, we need a thermal mapping to convert from the packing fraction into temperature. This thermal mapping is formulated based on the thermal expansion. The thermal expansion coefficient at ambient conditions is defined as
\begin{eqnarray}
\beta = \frac{1}{V}\left(\frac{\partial V}{\partial T}\right) = -\frac{1}{\Phi}\left(\frac{\partial \Phi}{\partial T}\right).
\label{eq:7}
\end{eqnarray}
where $V$ is a volume of material. A general solution of Eq. (\ref{eq:7}) is $\Phi = \Phi_0e^{-\beta(T-T_0)}$ with $\Phi_0 \approx 0.5$ being a characteristic volume fraction and $\beta \approx 12\times 10^{-4}$ $K^{-1}$ being a common value for many organic materials \cite{61,62,11,42,35}. This thermal mapping is also consistent with ref. \cite{40,41}. Since our ENCLE results find $\tau_\alpha = 100$s at $\Phi = \Phi_g \approx 0.611$, we determine $T_0$ by using the experimental glass transition temperature $T_g$, where $\tau_\alpha(T_g)$ = 100s. We have
\begin{eqnarray}
T_0 = T_g+\frac{1}{\beta}\ln\left(\frac{\Phi_g}{\Phi_0}. \right)
\label{eq:7-1}
\end{eqnarray}

At elevated pressures, the free volume is reduced and the above density-to-temperature conversion (thermal mapping) has to change. To propose a new thermal mapping including pressure effects, we use three assumptions: (i) the deformation obeys linear elasticity, (ii) the size and number of particles in hard-sphere fluids remain unchanged during compression, and (iii) the thermal expansion coefficient $\beta$ is weakly sensitive to pressure. In a prior work \cite{30}, $\beta$ of a triphenylchloromethane/o-terphenyl mixture decreases from $7.8\times 10^{-4}$ $K^{-1}$ to $6.9\times 10^{-4}$ $K^{-1}$ as increasing pressure from 0.1 MPa to 800 MPa. Another experimental study \cite{31} reveals a reduction of the thermal expansion coefficient of poly(vinyl acetate) from $7.1\times 10^{-4}$ $K^{-1}$ to $5.9\times 10^{-4}$ $K^{-1}$ in the same range of applied pressure. Since the variations are relatively small, our assumption (iii) seems reasonable for zeroth order approximation but the accuracy depends on materials.

According to Hooke's law, a relative change in length is
\begin{eqnarray}
\frac{\Delta L}{L} = \frac{P}{E},
\label{eq:9-1}
\end{eqnarray}
where $P$ is an external pressure, $E$ is an "effective" Young modulus, $L$ is the uncompressed length, and $\Delta L$ is the change in length under pressure. From this, the volume fraction is
\begin{eqnarray}
\Phi(P) &=& \frac{N\pi d^3}{6V(P)} = \frac{N\pi d^3}{6V(P=0)(1-\Delta L/L)}\nonumber\\
&\approx& \Phi(P=0)\left(1+\frac{\Delta L}{L} \right) \nonumber\\
&=& \Phi_0e^{-\beta(T-T_0)}\left(1+\frac{P}{E}\right),
\label{eq:9}
\end{eqnarray}
where $N$ is the number of hard-sphere particles, $V(P)$ and $V(P=0)$ are the volume at pressure $P$ and atmospheric pressure, respectively. The low-frequency (low deformation rate) or static modulus obviously obeys the Hooke's law and is very small in liquid state. At high frequency, the modulus (given by Zwanzig-Mountain formula \cite{43,40}) cannot be used in the Hooke's law, unless the stress-strain relation is measured at extremely high strain rates. Both the low-frequency and high-frequency moduli slightly change with temperature below $T_g$ but significantly decrease with temperature above $T_g$ \cite{40}. For strong glass-forming liquids, effects of pressure and temperature on the change of the elastic modulus are minor \cite{41,33,34}. Our effective modulus is an average quantity over a wide range of temperature and pressure considered in experiments. An increase of $P$ and a decrease of $T$ is roughly equivalent in the aspect of reducing free volume. Thus, a material is characterized by only one value of the modulus. There is no pressure and temperature independence of $E$. The definition of $E$ is supposed to keep the validity of the Hooke's law in our calculations. The modulus can be representative for the elastic stiffness of a given material. 


From the theoretical $\tau_\alpha(T,P)$, we determine the dynamic fragility
\begin{eqnarray}
m = \left. \frac{\partial\log_{10}(\tau_\alpha)}{\partial(T_g/T)}\right |_{T=T_g}.
\label{eq:fragility}
\end{eqnarray} 
This physical quantity is used to classify amorphous materials into "strong" ($m \leq 30$) or "fragile" ($m \geq 100$) materials. For $100 > m >30$, materials are intermediate glass formers.

\section{Results and discussions}
\begin{figure*}[htp]
\center
\includegraphics[width=8cm]{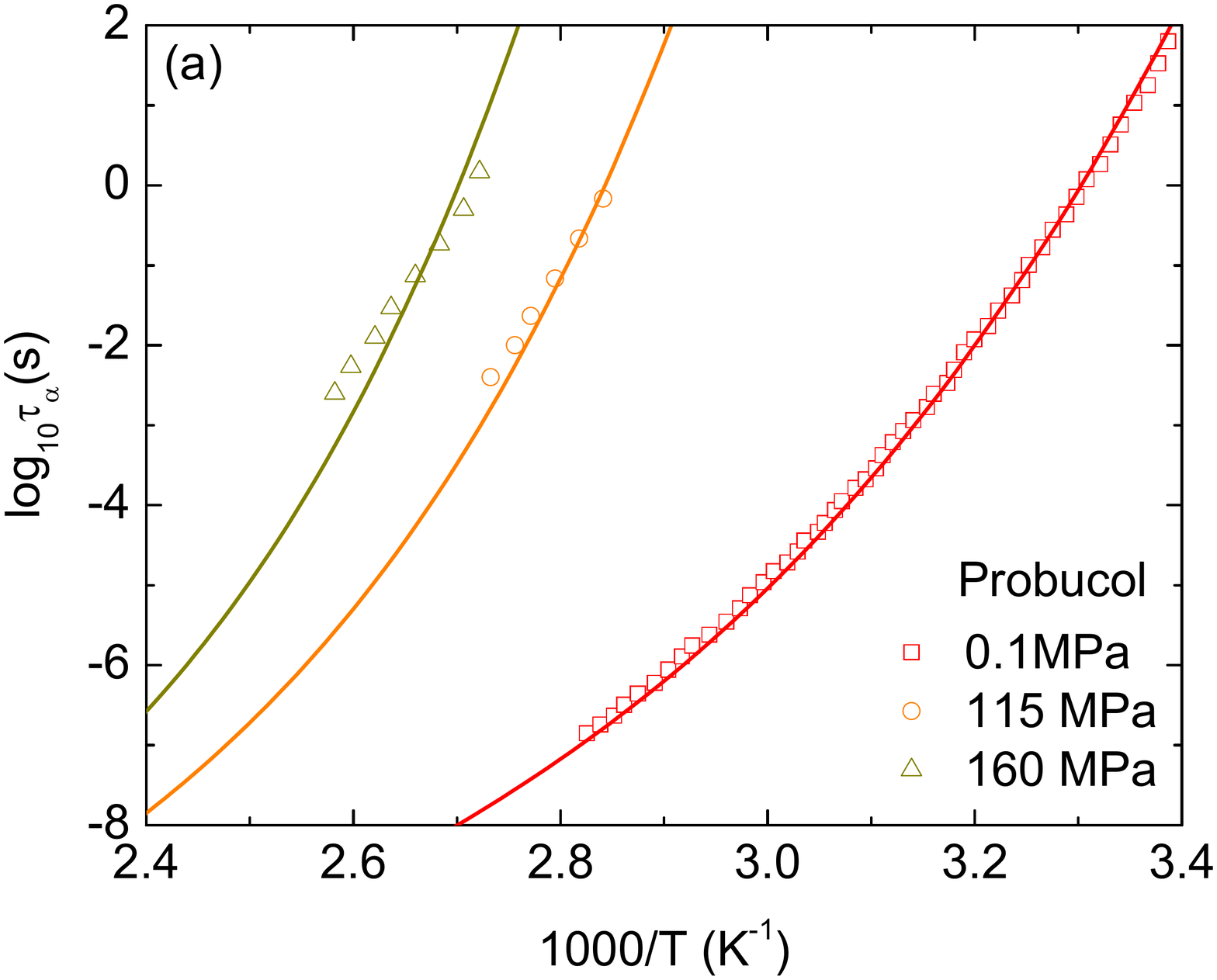}
\includegraphics[width=8cm]{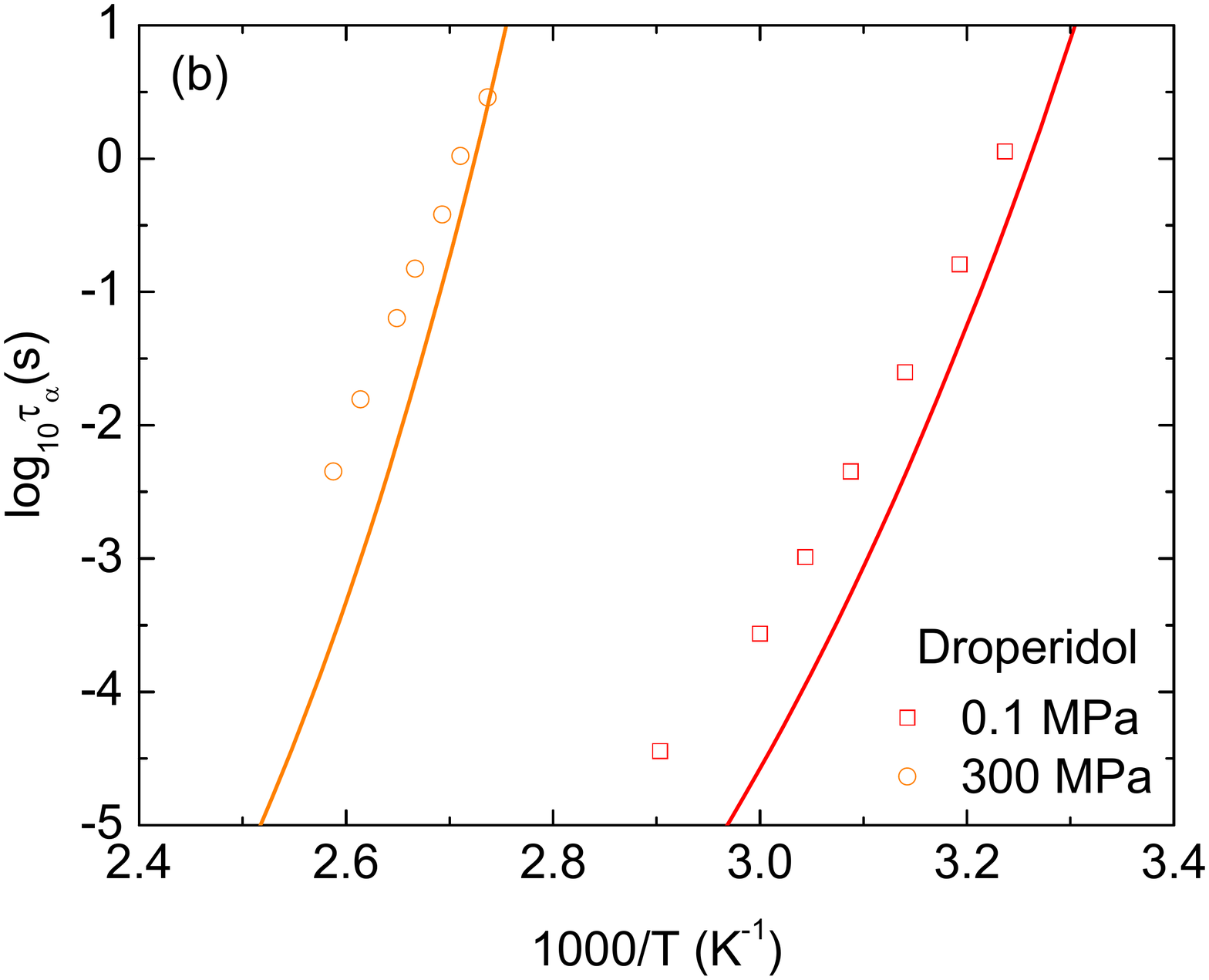}
\includegraphics[width=8cm]{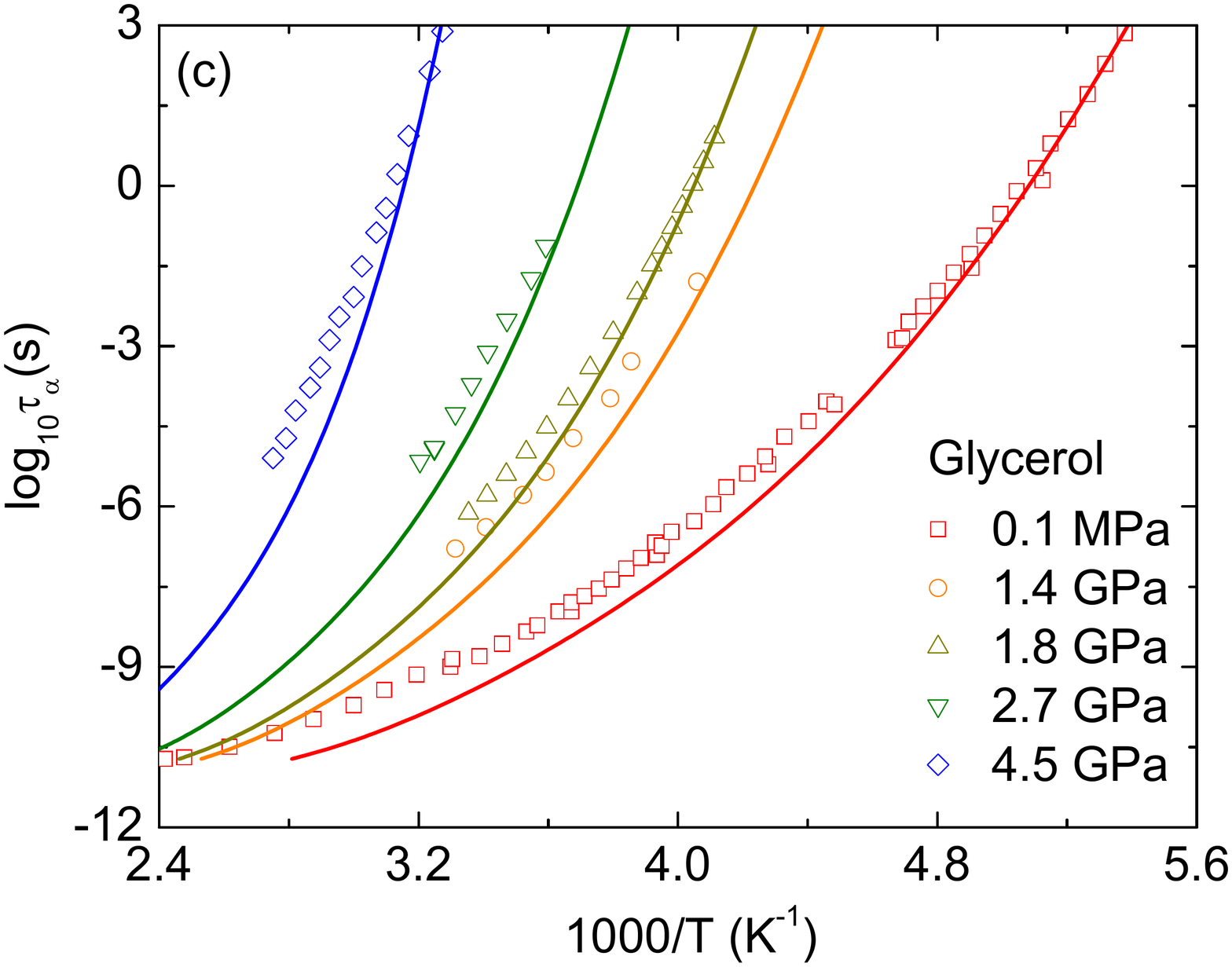}
\includegraphics[width=8cm]{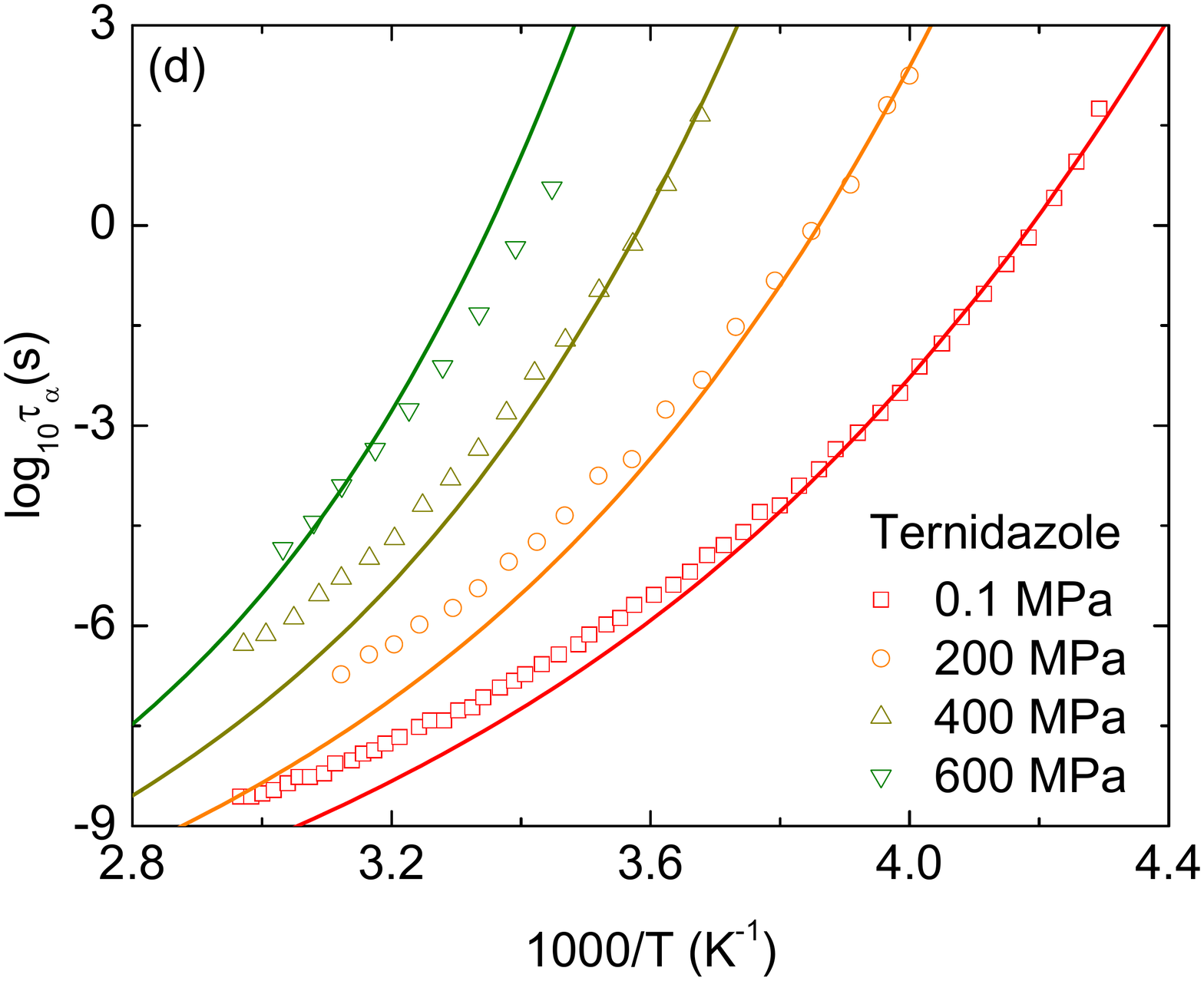}
\caption{\label{fig:1}(Color online) The temperature dependence of $\tau_\alpha$ of (a) probucol, (b) droperidol, (c) glycerol, and (d) temidazole at different pressures. Open points are experimental data in ref. \cite{70,71,72,24,25} and solid curves correspond to ECNLE calculations.}
\end{figure*}

\begin{figure*}[htp]
\includegraphics[width=12cm]{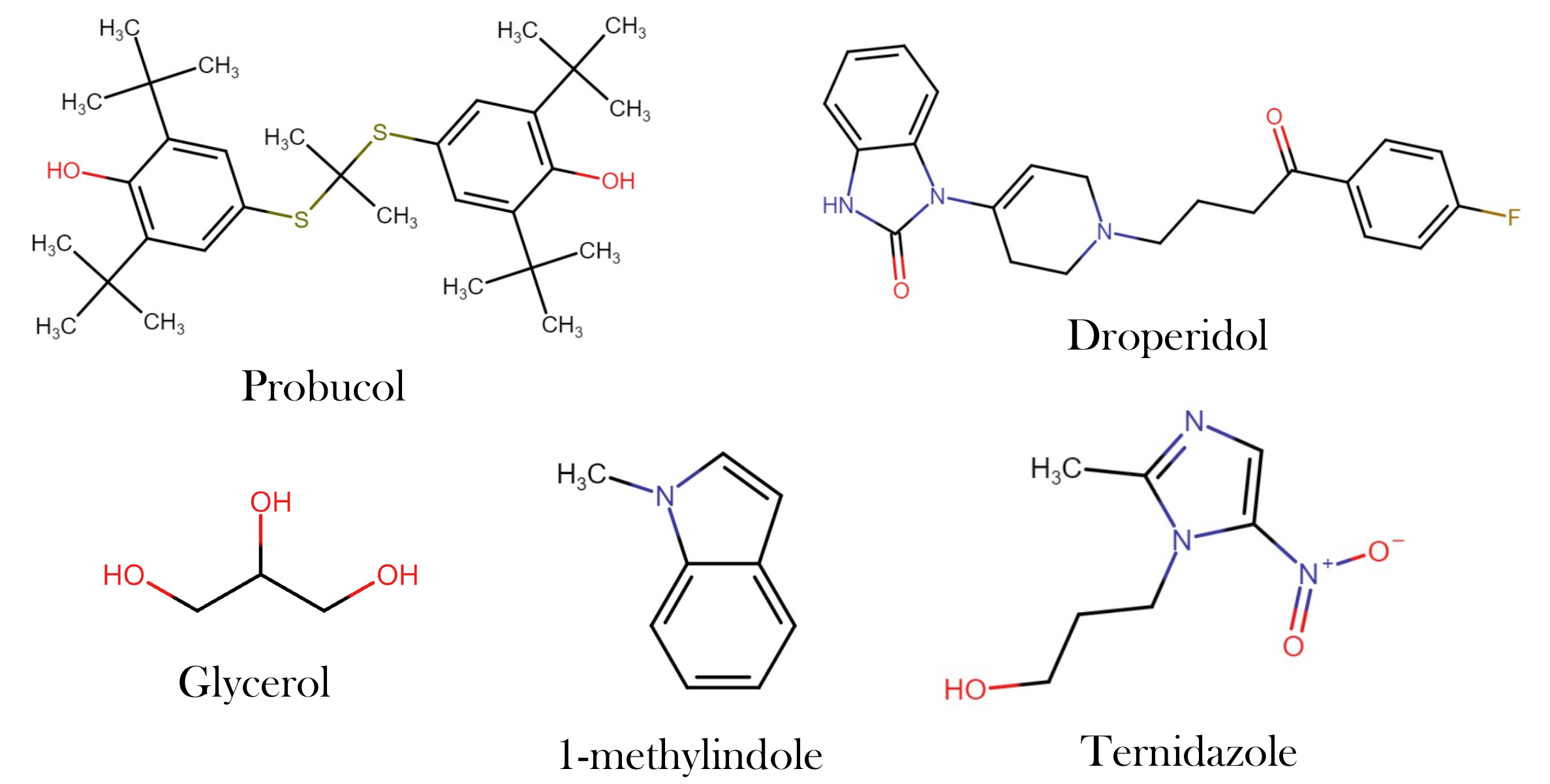}
\caption{\label{fig:0}(Color online) Chemical structures of investigated materials.}
\end{figure*}

Figure \ref{fig:1} shows the logarithm of theoretical and experimental structural relaxation time of probucol, droperidol, glycerol, and temidazole as a function of temperature at different pressures. We use eq. (\ref{eq:9}) for $\Phi$ in eqs. (\ref{eq:6}) and (\ref{eq:8}) to calculate $\tau_\alpha(T,P)$. Chemical structures of these materials are shown in Figure \ref{fig:0}. The Young's modulus $E$ of each material is chosen to obtain the best fit between theory and experiments at an elevated pressure $P^*$. Then, we can predict $\tau_\alpha(T,P)$ at pressures larger and smaller than $P^*$ without any additional parameter. Theoretical curves are closer to experimental data points at low pressure than at high pressure since the validity of assumptions of $\beta(P)\approx\beta(P=0)$ is reduced at high pressure. In Eq. (\ref{eq:6}), we assume a universal correlation between local and collective dynamics for all materials. However, our calculations for droperidol do not give a quantitative good description for experiment. It suggests that the correlation is non-universal and this result is consistent with prior works \cite{61,35,11}. Generically, ECNLE theory agrees with experimental data for other materials. Since our ECNLE approach is based on the hard sphere model, it works well for nonpolar van der Waals (vdW) liquids \cite{10}. The presence of hydrogen bonding interactions in materials causes more chemical and conformational complexities, and reduces accuracy of theoretical predictions. Thus, three main reasons of a remarkable deviation at high pressures are: (i) biological/chemical conformation can be altered, (ii) hydrogen bonding effects are significantly enhanced due to an increase of hydrogen bond density, and (iii) effects of compression on thermodynamic quantities are ignored in our calculations.  

\begin{figure}[htp]
\center
\includegraphics[width=8.5cm]{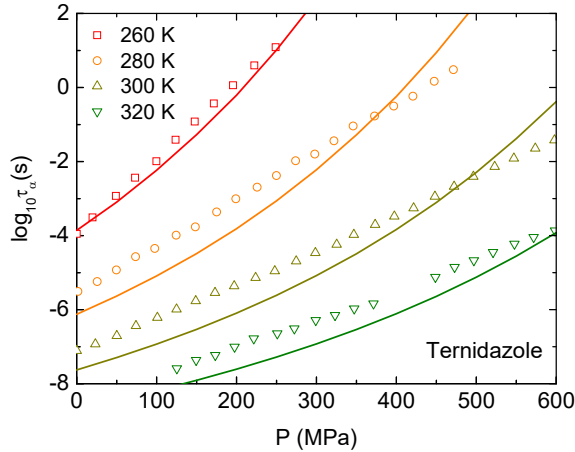}
\caption{\label{fig:2}(Color online) The pressure dependence of $\log_{10}\tau_\alpha$ of temidazole at isothermal conditions. Open points are experimental data in ref. \cite{71} and solid curves correspond to ECNLE calculations.}
\end{figure}

The agreement between theory and experiments partially validates our assumptions in this approach. Thus, we can calculate the effective Young modulus of studied materials and report results on Table \ref{table:1}. The values of probucol, droperidol, and temidazole are in the same order of magnitude as many drugs, polymers, and thermal liquids \cite{14} measured by the dynamic mechanical analyzer at room temperature and a frequency of 0.17 Hz. For glycerol, our elastic modulus is two times larger than the experimental counterpart deduced from ref. \cite{15}. This finding suggests that theoretical results may be overpredicted by a factor of 2. The difference is expected due to simplification of model and definition of the elastic modulus. It needs to be tested by future simulations and experiments. From these obtained elastic moduli, we determine the pressure dependence of $\tau_\alpha$ of temidazole at fixed temperatures and present numerical results in Figure \ref{fig:2}. Remarkably, there is a slight difference between theory and experiment over a wide range of pressure. 



\begin{figure}[htp]
\center
\includegraphics[width=8.5cm]{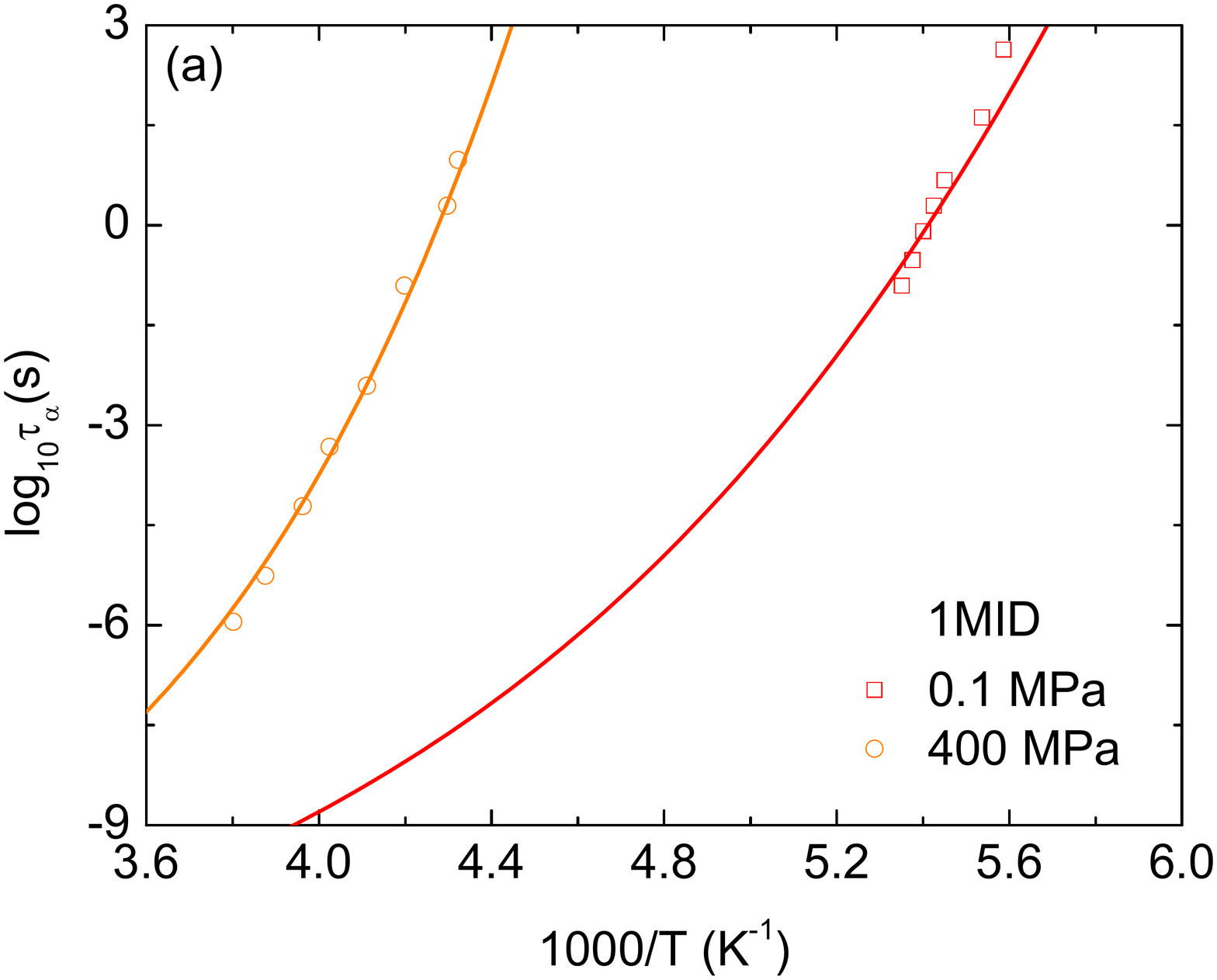}
\includegraphics[width=8.5cm]{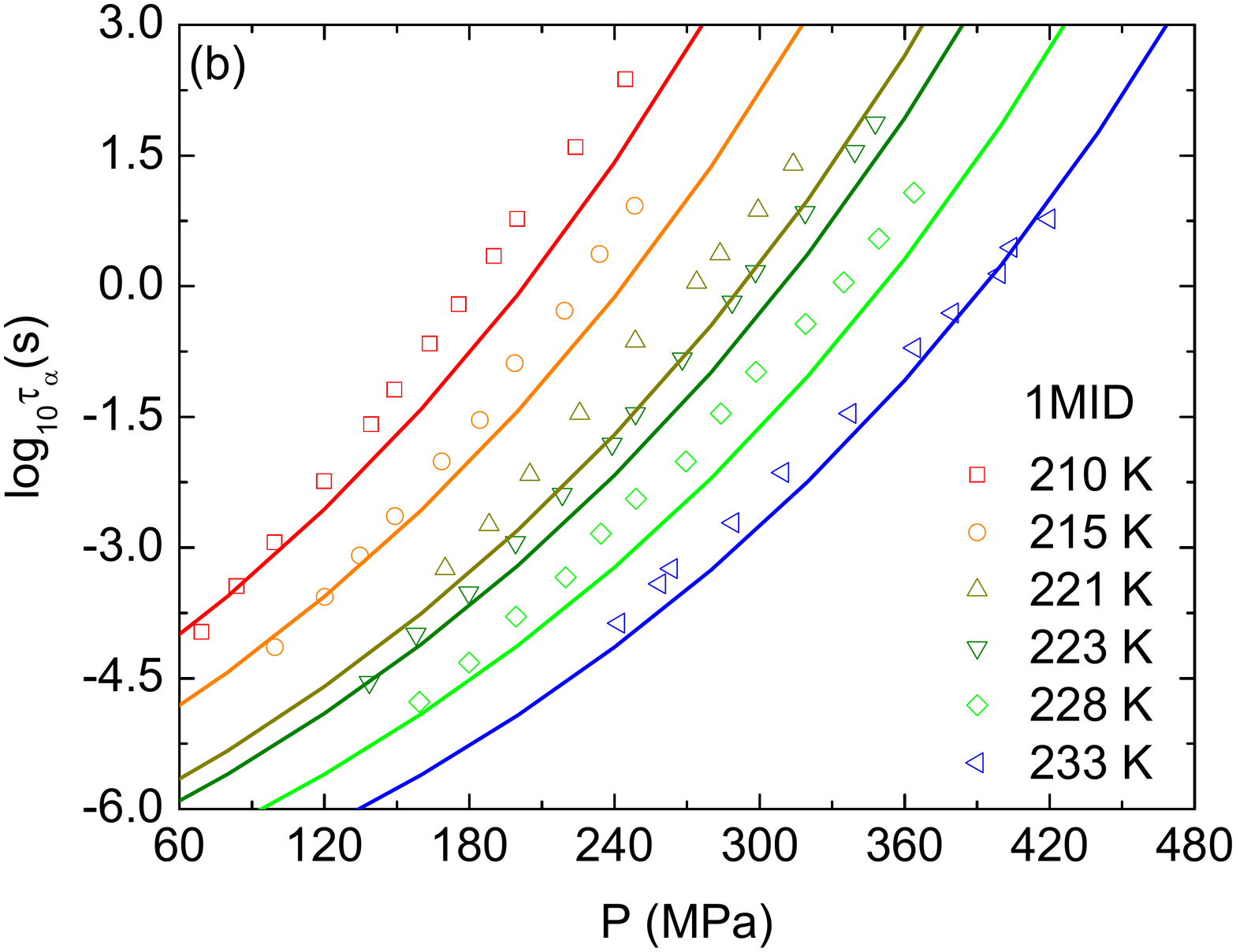}
\caption{\label{fig:3}(Color online) (a) The temperature dependence of $\log_{10}\tau_\alpha$ at P = 0.1 MPa and 400 MPa, and (b) the pressure dependence of $\log_{10}\tau_\alpha$ of 1MID at different isobaric processes. Open points are experimental data in ref. \cite{70} and solid curves correspond to ECNLE calculations.}
\end{figure}

Numerical results allow us to determine theoretically the glass transition temperatures and dynamic fragilities, and contrast these predictions with experimental values in Table \ref{table:1}. The experimental and theoretical $T_g$s are relatively close together as a consequence of good agreement seen in Figure \ref{fig:1}. Note that, in ref. \cite{10}, the ECNLE theory provided a qualitative description for $T_g(P)$ and $m(P)$ of orthoterphenyl, toluene, glycerol, and salol. Authors used the equality between experimental and hard-sphere dimensionless compressibility to formulate a different thermal mapping. Although their approach does not have any fit parameter and the variations for vdW liquids are correct, numerical results are overpredicted. 

Our predicted fragilities indicate that all four hydrogen-bonding glass-forming liquids become more fragile with compression. This conclusion is completely consistent with glycerol's data in this work and prior study \cite{24,25} but behaves oppositely with probucol, droperidol, and other materials \cite{68}. There are three possible reasons for this incorrect direction of $m(P)$. Firstly, a hard sphere model is insufficiently good to investigate hydrogen-bonding liquids. However, ECNLE predictions of ref. \cite{10} using the hard-sphere model and compressibility-based thermal mapping exhibit a decrease of $m(P)$ with pressure up to 75 MPa. This density-to-temperature conversion not only has less assumptions, but also captures more information at elevated pressures. Secondly, more complexities of molecular conformation and volume occur as increasing pressure. An applied pressure has a strong influence on steric effects, which cause a pressure-dependent correlation between local and collective dynamics. Both our work and ref. \cite{10} ignore this issue. Thirdly, we suspect that a main problem of our predicted fragility lie in our thermal mapping, particularly assumptions of the pressure independence of $\beta$ and $E$. A small increase in these quantities dramatically changes the slope of relaxation time with temperature in isobaric processes, which is tightly related to $m(P)$. This problem is under study.

To investigate roles of molecular shape anisotropy on the glass transition, we consider a liquid of 1-methylindole (1MID) molecules, which are very rigid and planar. The in-plane translational motion of molecules is favored rather than the out-of-plane motion because of a larger free volume necessary for out-of-plane movements. Thus, the jump distance is altered in the spherical mean-field average. To handle this problem, we adopt a physical idea in ref. \cite{36, 37} and introduce a parameter to scale the jump distance by a factor of 1.5, $\Delta r \equiv 1.5\Delta r$, to obtain the best accordance between theoretical and experimental alpha relaxation time at atmospheric pressure. This scaling increases contributions of the collective elastic barrier to the glass transition and leads to $\tau_\alpha(\Phi_g=0.5828)=100$ s. Inserting the new $\Phi_g$ into Eqs. (\ref{eq:7-1}) and (\ref{eq:9}) gives the temperature dependence of $\tau_\alpha$ at a certain pressure. Then, we calculate the structural relaxation time of 1MID at isothermal and isobaric processes, and show numerical results in Figure \ref{fig:3} with experiments. Theoretical curves and experimental data in Figure \ref{fig:3} exhibit better consistency than those in Figure \ref{fig:2}. Moreover, the predicted and experimental $T_g$ are also close to each other at elevated pressure ($P =400$ MPa) as shown in Table \ref{table:1}.

\begin{table}[htp]
\caption{The glass transition temperatures in Kelvin and dynamic fragility predicted by ECNLE theory (th) and their corresponding experimental values (expt) in ref. \cite{70,71,72,25}. The units of effective Young modulus and pressure are GPa.}
\centering 
\begin{tabular}{|c | c| c| c| c| c| c|} 
\hline
Material & $P$& $T_g$ (th) & $T_g$ (epxt) & $m$ (th) & $m$ (epxt)& $E$\\ [0.5ex] 
\hline 
probucol & $10^{-4}$ & 294.6 & 294.6 & 83.9 & 83 & 1.9\\ 
& 0.115 & 343.9 & 342 & 97.3 & 68.5 & \\ 
& 0.16 & 362.5 & 355.6 & 102.3 & 66& \\ 
[0.5ex] 
\hline
droperidol& $10^{-4}$ & 298.4 & 298.4  & 85.3 & 75 & 4\\
& 0.3 & 358.5 & 356.2 & 103 & 68.4& \\
\hline
glycerol& $10^{-4}$ & 190 & 190 & 52.5 & 52.2 & 29\\
& 1.8 & 242 & 239.2 & 63.7 & 67.6 & \\
\hline
temidazole& $10^{-4}$ & 231.9 & 231.9 & 65.3 & 76.8 & 8.1\\
& 0.2 & 251 & 251.2 & 71.3 & 73.3 & \\
& 0.4 & 271 & 270.8 & 76.8 & 71.0 & \\
& 0.6 & 290 & 284 & 82.8 & 66.1 & \\
\hline
1MID& $10^{-4}$ & 180 & 180 & 63.5 & & 6.6\\
& 0.4 & 227 & 228 & 81.1 & & \\
[0.5ex]
\hline
\end{tabular}
\label{table:1} 
\end{table}
\section{Conclusions}
In conclusion, we have extended the ECNLE theory to investigate pressure effects on the glass transition of probucol, droperidol, glycerol, temidazole, and 1MID. We employ a hard-sphere model to describe glass-forming liquids in the same manner as the original version \cite{2,7,10,6,35,42,11,61,62}, but propose a simple and new thermal mapping to convert from a packing fraction of the fluid to temperature at a given pressure. The formulation of our thermal mapping is based on the thermal expansion and linear elastic compression, which is related to the effective elastic modulus of investigated materials. This modulus is determined in an average manner over temperature and pressure to quantify the elastic stiffness of a amorphous material. From this, we intercorrelate density, temperature, and pressure. To simplify calculations, we assume that the thermal expansion coefficient is weakly dependent on pressure. By adjusting the modulus to have the best agreement between experimental and theoretical relaxation time at a certain pressure, we determine this elastic modulus. Interestingly, theory quantitatively agree with experiment over 14 decades in time and a wide range of pressure (up to 4.5 GPa). For the first time, our findings indicate the broadband dielectric spectroscopy can be used to determine the effective modulus. 

Although our calculations have successfully described the temperature and pressure dependence of structural relaxation time, a decrease of the predicted dynamic fragility with increasing compression is opposite to experiments. Three reasons leading to an incorrect prediction of $m(P)$ are: (i) inaccuracy of a hard-sphere model in studying the glassy dynamics of hydrogen-bonding glass formers, (ii) pressure-induced chemical/biological complexities, and (iii) underestimation of pressure sensitivity of $\beta$ and $E$. We suspect the last reason is a main issue. These are open problems for scientific community.

Our approach is similar to ref. \cite{10} except for the thermal mapping. This approach based on experimental dimensionless compressibility data predicts an increase of $T_g$ and a decrease of $m$ as increasing pressure (up to 75 MPa). These results qualitatively agree with experiments. The fact is that the thermal mapping of ref. \cite{10} uses many experimental results at high pressure. Thus, there is more information of the pressure dependence encoded in this mapping than our mapping. However, their theoretical $T_gs(P)$ quantitatively deviate from experimental results. The predicted structural relaxation times were not compared to the experimental counterparts. When a lack of experimental dimensionless-compressibility data occurs, this approach cannot be implemented. The validity of the thermal mapping in ref. \cite{10} may be limited within the pressure range of experimental data.

\begin{acknowledgments}
This research was funded by the Vietnam National Foundation for Science and Technology Development (NAFOSTED) under grant number 103.01-2019.318. 
\end{acknowledgments}

\end{document}